\documentclass[aps,prl,onecolumn,showpacs,preprintnumbers]{revtex4}%
\usepackage{eurosym}
\usepackage{amsmath}
\usepackage{dcolumn}
\usepackage{bm}
\usepackage{subfigure}
\usepackage{amsfonts}
\usepackage{amssymb}
\usepackage{makeidx}
\usepackage{epsfig}
\usepackage{graphicx}%
\begin{document}
\title{Asymptotic orderings and approximations of the Master kinetic equation\\for large hard spheres systems}
\author{Massimo Tessarotto}
\affiliation{Department of Mathematics and Geosciences, University of Trieste, Via Valerio
12/1, 34127 Trieste, Italy}
\affiliation{Institute of Physics, Faculty of Philosophy and Science, Silesian University
in Opava, Bezru\v{c}ovo n\'{a}m.13, CZ-74601 Opava, Czech Republic}
\author{Claudio Asci}
\affiliation{Department of Mathematics and Geosciences, University of Trieste, Via Valerio
12/1, 34127 Trieste, Italy}
\date{\today }

\begin{abstract}
In this paper the problem is posed of determining the physically-meaningful
asymptotic orderings holding for the statistical description of a large
$N-$body system of hard spheres,\textit{ i.e.,} formed by $N\equiv\frac
{1}{\varepsilon}$\ $\gg1$ \ particles, which are allowed to undergo
instantaneous and purely elastic unary, binary or multiple collisions.
Starting point is the axiomatic treatment recently developed [Tessarotto
\textit{et al}., 2013-2016] and the related discovery of an exact kinetic
equation realized by Master equation which advances in time the $1-$body
probability density function (PDF) for such a system. As shown in the paper
the task involves introducing appropriate asymptotic orderings in terms\ of
$\varepsilon$ for all the physically-relevant parameters. The goal is that of
identifying the relevant physically-meaningful asymptotic approximations
applicable for the Master kinetic equation, together with their possible
relationships with the Boltzmann and Enskog kinetic equations, and holding in
appropriate asymptotic regimes. These correspond either to dilute or dense
systems and are formed either by small--size or finite-size identical hard
spheres, the distinction between the various cases depending on suitable
asymptotic orderings in terms\ of $\varepsilon.$

\end{abstract}

\pacs{05.20.-y, 05.20.Dd, 05.20.Jj, 51.10.+y}
\keywords{kinetic theory, classical statistical mechanics, Boltzmann equation, H-theorem}\maketitle



\section{1 - INTRODUCTION}

In a series of papers \cite{noi1,noi2,noi3,noi4,noi5,noi6,noi7} (see also
Refs. \cite{noi8,noi9}) a new kinetic equation has been established for hard
sphere systems subject to elastic instantaneous collisions, denoted as Master
kinetic equation. Its basic features are that, unlike the Boltzmann and Enskog
kinetic equations \cite{Boltzmann1972,Enskog,CHAPMA-COWLING}, the same
equation and its corresponding Master collision operator are exact,\textit{
i.e.,} they hold for an arbitrary$N-$body hard-sphere system $S_{N}$ which is
isolated, namely\textit{ }for which the number of particles ($N$) is constant,
while furthermore that its solutions are entropy-preserving \cite{noi6} and
globally defined \cite{noi7}. Concerning, in particular, the first feature
this means that in such a context $S_{N}$ is allowed to have an arbitrary
\emph{finite number} $N$ of \emph{finite-size }and\emph{ finite-mass }hard
spheres, namely each one characterized by finite diameter $\sigma>0$ and mass
$m>0.$ In addition, by assumption $S_{N}$ is immersed in a bounded
configuration domain $\Omega,$ subset of the Euclidean space $%
\mathbb{R}
^{3}$ which has a finite canonical measure $L_{o}^{3}\equiv\mu(\Omega)>0$
($L_{o}$ denoting a finite configuration-domain characteristic scale length)
and is endowed with a stationary and rigid boundary $\partial\Omega$. However,
the total volume occupied by hard-spheres cannot exceed the
configuration-space volume. Hence, the parameters $N,L_{o}$ and $\sigma$ must
necessarily satisfy the inequality
\begin{equation}
\Delta\equiv\frac{N\mu(\Phi)}{\mu(\Omega)}\equiv\frac{4\pi N\sigma^{3}}%
{3L_{o}^{3}}\leq1,\label{GLOBAL DILUTENESS}%
\end{equation}
with $\mu(\Phi)\equiv\frac{4\pi\sigma^{3}}{3}$ denoting the volume of a single
hard sphere \ and $\Delta$ the \emph{global diluteness parameter}. \ For such
an equation the particle correlations appearing through the $2-$body
probability density function (PDF) are exactly taken into account by means of
suitably-prescribed $1-$ and $2-$body occupation coefficients which are
position-dependent only$.$\ These peculiar features follow uniquely as a
consequence of the new approach to classical statistical mechanics developed
in Refs. \cite{noi1,noi2,noi3} and referred to as "\textit{ab initio}"
axiomatic approach. As shown in the same references (for a review see also
Ref. \cite{noi9}), this is based on the adoption of appropriate \emph{extended
functional setting} and physics-based \emph{modified collision boundary
conditions} (MCBC) \cite{noi1,noi3} which are prescribed in order to advance
in time across arbitrary (unary, binary or multiple) collision events the
$N-$body probability density function (PDF). The  physical origin of MCBC
follows from the requirement that the deterministic $N-$body Dirac delta is
included among the physically-admissible $N-$body PDF's for $S_{N}$
\cite{noi2}$.$ Its physical interpretation is intuitive \cite{noi9} being
viewed as the jump condition for the $N-$body PDF along the phase-space
Lagrangian trajectory $\left\{  \mathbf{x}(t)\right\}  $ for an ensemble of
$N$ tracer particles \cite{noi0} undergoing an arbitrary collision event$.$

Based on the discovery of the Master kinetic equation, a host of new
developments have opened up. These concern, first of all, the investigation of
novel conceptual aspects of the same equation (for an extended discussion see
Refs. \cite{noi4,noi5,noi6,noi7,noi8}). However, it goes almost without saying
that possible applications of the new equation are potentially ubiquitous.
Many of these applications typically concern \emph{large systems},\textit{
i.e.,} which are formed by a large number $N\equiv\frac{1}{\varepsilon}$%
\ $\gg1$ \ of particles.

The goal of this paper is that of identifying the relevant
physically-meaningful asymptotic approximations for the same kinetic equation
which may correspond to physically-relevant practical applications of the
theory. As shown below, this task involves the adoption of appropriate
asymptotic orderings in terms\ of $\varepsilon$ for all the
physically-relevant parameters. These include, besides the configuration-space
scale length $L_{o}$ and the hard-sphere diameter $\sigma$ also the
characteristic scale length $L_{\rho}$ which is associated with the spatial
variations of the $1-$body PDF which is prescribed so that
\begin{equation}
\frac{1}{L_{\rho}}=\sup\left\{  \left\vert \frac{\partial\ln\rho_{1}%
^{(N)}\left(  \mathbf{x}_{1},t\right)  }{\partial\mathbf{r}_{1}}\right\vert
,\text{ }\forall\left(  \mathbf{x}_{1},t\right)  \in\Gamma_{1(1)}\times
I\right\}  .
\end{equation}
Notice that hereon $\frac{\partial\ln\rho_{1}^{(N)}\left(  \mathbf{x}%
_{1},t\right)  }{\partial\mathbf{r}_{1}}$ is assumed to be bounded for all
$\left(  \mathbf{x}_{1},t\right)  $ spanning the extended $1-$body phase space
$\Gamma_{1(1)}\times I$ (see notations below). In the present paper the issues
are investigated which concern the classification of $N-$body systems which
are characterized respectively by suitable asymptotic orderings. More precisely:

\begin{itemize}
\item A) $\sigma-$\emph{ordering regime: }in which the diameter\emph{ }%
$\sigma$ of the hard spheres is either \emph{small-size} or \emph{finite
size}, in the sense that either $\sigma\sim O(\varepsilon^{\alpha})$ with
suitable $\alpha>0$ or $\sigma\sim O(\varepsilon^{0});$

\item B) $\Delta-$\emph{ordering regime: }in which the parameter\emph{
}$\Delta$ is either $\Delta\sim O(\varepsilon^{\beta})$ with $\beta>0$ or
$\Delta\sim O(\varepsilon^{0}),$ namely the hard-sphere system is
\emph{dilute} or \emph{dense;}
\end{itemize}

\noindent A type of alternate asymptotic ordering equivalent to B) is provided
also by:

\begin{itemize}
\item C) $K_{n}-$\emph{ordering regime: }in which a suitably-defined Knudsen
number $K_{n}$ (see below) may be either of order $O(\varepsilon^{\gamma}),$
being $\gamma$ a suitable positive, either vanishing or negative real number.
Accordingly the hard-sphere systems will be denoted as
\emph{weakly-collisional}, \emph{collisional} or \emph{strongly collisional.}
\end{itemize}

\noindent The topics indicated above include in particular:

\noindent\emph{ISSUE \#1:} the search of possible asymptotic (\textit{i.e.,}
approximate) kinetic equations holding in cases A, B and C;

\noindent\emph{ISSUE \#2:} the possible asymptotic evaluation of the $1-$ and
$2-$body occupation coefficients in the same cases.

\noindent The goal is also to display the possible relationship of the Master
kinetic equation with well-known kinetic equations,\textit{ i.e.,} the
Boltzmann and Enskog kinetic equations
\cite{Boltzmann1972,Enskog,CHAPMA-COWLING}. In particular, although in the
case of finite-size hard spheres the strict validity of both the Boltzmann and
Enskog equations "\textit{per se"} is ruled out \cite{noi3}, this concerns:

\noindent\emph{ISSUE \#3:} the determination of the asymptotic modifications
which enter the Boltzmann equation in the case the particle diameter $\sigma$
is suitably small.

\noindent\emph{ISSUE \#4:} the identification of the relevant asymptotic
parameter sub-domains in which the Enskog equation still applies, albeit in a
suitable approximate sense.\ 

\section{2 - Dimensionless representation of the Master kinetic equation}

The prerequisite for carrying out the tasks outlined above is setting the
Master kinetic equation in dimensionless form. To this end let us first notice
that by construction the $1-$body PDF $\rho_{1}^{(N)}\left(  \mathbf{x}%
_{1},t\right)  $ depends on the extended Newtonian state $\left(
\mathbf{x}_{1}\equiv\left\{  \mathbf{r}_{1},\mathbf{v}_{1}\right\}
_{1},t\right)  .$ In particular this means that:

1) the Newtonian position vector and velocity vectors $\mathbf{r}_{1}$ and
$\mathbf{v}_{1},$ which label the center of mass position and velocity of a
representative particle, span the Euclidean configuration and velocity spaces
$\Omega\subset%
\mathbb{R}
^{3}$ and $U_{1(1)\text{ }}\equiv%
\mathbb{R}
^{3},$ while $t$ belongs to the Galilean time axis $I\equiv%
\mathbb{R}
.$ As a consequence the Galilean structure of $\Omega\times I,$\textit{ i.e.,}
the Euclidean distance in $\Omega$ and the time-interval in $I$, remains
uniquely determined.

2) by construction $\rho_{1}^{(N)}\left(  \mathbf{x}_{1},t\right)  $ is a
scalar with respect to the group of Galilei transformation which preserves the
Galilean structure of the set $\Omega\times I$.

Next, let us introduce the \emph{characteristic scale length }
\begin{equation}
L\equiv\min\left\{  L_{o},L_{\rho}\right\}  ,\label{definition of L}%
\end{equation}
and a suitable constant \emph{characteristic time scale }$\tau$ (whose
definition remains in principle arbitrary). Then all the Newtonian variables
$\left(  \mathbf{x}_{1}\equiv\left\{  \mathbf{r}_{1},\mathbf{v}_{1}\right\}
,t\right)  $ can be conveniently replaced with the corresponding dimensionless
quantities $\overline{\mathbf{r}}_{1}=\frac{1}{L}\mathbf{r}_{1},\overline
{\mathbf{v}}_{1}=\frac{\tau}{L}\mathbf{v}_{1}$ and $\overline{t}=\frac{1}%
{\tau}t.$ This implies that the phase-space volume element must transform as
$d\mathbf{r}_{1}d\mathbf{v}_{1}=\frac{L^{6}}{\tau^{3}}d\overline{\mathbf{r}%
}_{1}d\overline{\mathbf{v}}_{1}$, while, in order to warrant the conservation
of probability $d\mathbf{r}_{1}d\mathbf{v}_{1}\rho_{1}^{(N)}(\mathbf{r}%
_{1},\mathbf{v}_{1},t)=d\overline{\mathbf{r}}_{1}d\overline{\mathbf{v}}%
_{1}\overline{\rho}_{1}^{(N)}$, the dimensionless form, of the PDF
$\overline{\rho}_{1}^{(N)}$ must be identified with $\overline{\rho}_{1}%
^{(N)}=\frac{L^{6}}{\tau^{3}}\rho_{1}^{(N)}(\mathbf{r}_{1},\mathbf{v}_{1},t)$.
Notice here, however, that to preserve the scalar property of the transformed
PDF $\overline{\rho}_{1}^{(N)}$ with respect to the Galilei group the latter
should depend explicitly on the extended Newtonian state\textbf{ }%
($\mathbf{r}_{1},\mathbf{v}_{1},\overline{t}$) rather then the transformed
state ($\overline{\mathbf{r}}_{1},\overline{\mathbf{v}}_{1},\overline{t}$). In
fact due to their arbitrariness, the parameters $L$ and $\tau$ change the
Galilei structure of space-time,\textit{ i.e.}, the Euclidean distance and the
time interval.\textbf{ }Hence $\overline{\rho}_{1}^{(N)}$ still depends, as
$\rho_{1}^{(N)}(\mathbf{r}_{1},\mathbf{v}_{1},t),$ on the same extended state
$\left(  \mathbf{x}_{1}\equiv\left\{  \mathbf{r}_{1},\mathbf{v}_{1}\right\}
_{1},t\right)  ,$ and therefore is of the form $\overline{\rho}_{1}%
^{(N)}\equiv\overline{\rho}_{1}^{(N)}(\mathbf{r}_{1},\mathbf{v}_{1},t)$. In
terms of such a prescription the Master kinetic equation (see Ref.\cite{noi3})
can therefore be formally represented in the dimensionless form%
\begin{equation}
\overline{L}_{1}\overline{\rho}_{1}^{(N)}=\overline{C}_{1}(\left.
\overline{\rho}_{1}^{(N)}\right\vert \overline{\rho}_{1}^{(N)}%
),\label{MASTER-EQ}%
\end{equation}
with $\overline{L}_{1}$ and $\overline{C}_{1}(\left.  \overline{\rho}%
_{1}^{(N)}\right\vert \overline{\rho}_{1}^{(N)})$ denoting respectively the
free-streaming and Master collision operators. Both are cast in the
dimensionless representation,\textit{ i.e.}, so that $\overline{L}_{1}%
\equiv\frac{\partial}{\partial\overline{t}}+\overline{\mathbf{v}}_{1}%
\cdot\frac{\partial}{\partial\overline{\mathbf{r}}_{1}}$ and
\begin{gather}
\overline{C}_{1}(\left.  \overline{\rho}_{1}^{(N)}\right\vert \overline{\rho
}_{1}^{(N)})=K_{n}\int\limits_{U_{1(2)}}d\overline{\mathbf{v}}_{2}\int
^{(-)}d\Sigma_{21}.\nonumber\\
\left[  \overline{\rho}_{2}^{(N)}(\mathbf{x}^{(2)(+)},\overline{t}%
)-\overline{\rho}_{2}^{(N)}(\mathbf{x}^{(2)},\overline{t})\right]  \left\vert
\overline{\mathbf{v}}_{21}\cdot\mathbf{n}_{21}\right\vert \overline{\Theta
}^{\ast}.\label{MASTER-coll}%
\end{gather}
In addition,\ in the Eq. (\ref{MASTER-EQ}) $K_{n}$ identifies the
\emph{Knudsen number}
\begin{equation}
K_{n}\equiv\frac{\left(  N-1\right)  \sigma}{L^{2}}^{2},\label{KNUDSEN-N}%
\end{equation}
while $\overline{\rho}_{2}^{(N)}(\overline{\mathbf{x}}^{(2)},\overline{t})$ is
the dimensionless $2-$ body PDF given by
\begin{gather}
\overline{\rho}_{2}^{(N)}(\mathbf{r}_{1},\mathbf{v}_{1},\mathbf{r}%
_{2},\mathbf{v}_{2},t)\equiv f(\mathbf{r}_{1},\mathbf{r}_{2},t)\times
\nonumber\\
\overline{\rho}_{1}^{(N)}(\mathbf{r}_{1},\mathbf{v}_{1},t)\overline{\rho}%
_{1}^{(N)}(\mathbf{r}_{2},\mathbf{v}_{2},t),
\end{gather}
with $f(\mathbf{r}_{1},\mathbf{r}_{2},t)$ denoting the position-dependent
dimensionless weight-factor
\begin{equation}
f(\mathbf{r}_{1},\mathbf{r}_{2},t)\equiv\frac{k_{2}^{(N)}(\mathbf{r}%
_{1},\mathbf{r}_{2},t)}{k_{1}^{(N)}(\mathbf{r}_{1},t)k_{1}^{(N)}%
(\mathbf{r}_{2},t)}.\label{OCCUPATION COEFFICIENTS TERM}%
\end{equation}
The remaining notations are standard \cite{noi3}. Thus $U_{1(k)}\equiv%
\mathbb{R}
^{3}$ is the $1-$body velocity space for the $k-$th particle$,$ the symbol
$\int^{(-)}d\Sigma_{21}$ denotes integration on the subset of the solid angle
of incoming particles namely for which $\overline{\mathbf{v}}_{12}%
\cdot\mathbf{n}_{12}<0,$ $\overline{\Theta}^{\ast}$ denotes\emph{\ }%
$\overline{\Theta}^{\ast}\equiv\overline{\Theta}\left(  \left\vert
\mathbf{r}_{2}-\frac{\sigma}{2}\mathbf{n}_{2}\right\vert -\frac{\sigma}%
{2}\right)  ,$ with $\overline{\Theta}(x)$ being the strong theta function,
while everywhere in the operator $\overline{C}_{1}(\left.  \overline{\rho}%
_{1}^{(N)}\right\vert \overline{\rho}_{1}^{(N)}),$ $\mathbf{r}_{2}$ is
identified by construction with $\mathbf{r}_{2}=\mathbf{r}_{1}+\sigma
\mathbf{n}_{21}.$ Furthermore $k_{1}^{(N)}(\mathbf{r}_{1},t),k_{1}%
^{(N)}(\mathbf{r}_{2},t)$ and $k_{2}^{(N)}(\mathbf{r}_{1},\mathbf{r}_{2},t)$
identify the dimensionless $1-$ and $2-$body occupation coefficients, whose
definitions in terms of the dimensionless $1-$body PDF are respectively:
\begin{equation}
\left\{
\begin{array}
[c]{c}%
k_{1}^{(N)}(\mathbf{r}_{1},t)=\int\limits_{\overline{\Gamma}_{1(2)}}%
d\overline{\mathbf{x}}_{2}\frac{\overline{\rho}_{1}^{(N)}(\mathbf{x}_{2}%
,t)}{k_{1}^{(N)}(\mathbf{r}_{2},t)}k_{2}^{(N)}(\mathbf{r}_{1},\mathbf{r}%
_{2},t),\\
k_{2}^{(N)}(\mathbf{r}_{1},\mathbf{r}_{2},t)=\int\limits_{\overline{\Gamma
}_{1(3)}}d\overline{\mathbf{x}}_{3}\frac{\overline{\rho}_{1}^{(N)}%
(\mathbf{x}_{3},t)}{k_{1}^{(N)}(\mathbf{r}_{3},t)}...\int\limits_{\overline
{\Gamma}_{1(N)}}d\overline{\mathbf{x}}_{N}\frac{\overline{\rho}_{1}%
^{(N)}(\mathbf{x}_{N},t)}{k_{1}^{(N)}(\mathbf{r}_{N},t)},
\end{array}
\right.  \label{OCC-1}%
\end{equation}
where once the position of particle $1$ is assumed prescribed, $\overline
{\Gamma}_{1(2)},\overline{\Gamma}_{1(3)},...\overline{\Gamma}_{1(N)}$ are the
admissible subsets of the $1-$body phase spaces of particles $2,3,..,N,$
$\Gamma_{1(2)},\Gamma_{1(3)},...\Gamma_{1(N)}$ obtained by subtracting
respectively the forbidden subsets $\Phi_{12}$ (from $\Gamma_{1(2)}$),
$\Phi_{13}\cup\Phi_{23}$ (from $\Gamma_{1(3)}$), $%
{\displaystyle\bigcup\limits_{i=2,N-1}}
\Phi_{iN}$ (from $\Gamma_{1(N)}$).

A remark is in order concerning the comparison with the analogous
dimensionless representation introduced originally by Grad \cite{Grad} for the
BBGKY hierarchy and the Boltzmann equation in particular (see also
Ref.\cite{Cercignani1975}). The basic departure of Eq. (\ref{MASTER-EQ}) with
respect to the latter equation lies of course in the different realization of
the collision operator. However, another major\ difference arises because of
the explicit introduction of the characteristic scale length $L$ in the
definition of the Knudsen number given in Eq. (\ref{KNUDSEN-N})$.$ Such a
choice is actually required in order to permit the distinction between
different asymptotic ordering regimes (see next Section), while, in contrast,
Grad's approach dealt only with the so-called Boltzmann-Grad limit. Indeed, as
shown below, it was actually appropriate for the treatment of the so-called
dilute-gas asymptotic ordering only, namely for the case in which both the
scale length $L_{o}$ and $L$ are considered of order $O(\varepsilon^{0})$. 

\section{3 - Classification of the asymptotic ordering regimes}

In this section the relevant asymptotic orderings are determined which are
applicable in the case of large hard-sphere systems,\textit{ i.e.,} for which
$N\equiv\frac{1}{\varepsilon}\gg1$ and subject to the validity of the volume
constraint inequality (\ref{GLOBAL DILUTENESS})$.$ Based on the prescription
of the Knudsen number (\ref{KNUDSEN-N}) we are now able to show that the
classification in terms of $K_{n}$\emph{ } pointed out above corresponds to
suitably prescribe the magnitude of the ratio dimensionless $\sigma/L.$ More
precisely \emph{weakly-collisional}, \emph{collisional} or \emph{strongly
collisional asymptotic regimes }are obtained requiring that $\sigma/L$ be of
order $O(\varepsilon^{\frac{\gamma-1}{2}})$ with $\gamma$ being respectively
$\gamma<0,$ $\gamma=0$ and $\gamma>0.$ For definiteness let us initially focus
on the collisional $K_{n}-$ordering regime. In this case the following two
possible \emph{dilute-gas ordering regimes} (3A and 3B) can be distinguished.

\subsection{3A - Dilute-gas small-size $\sigma-$ordering regime}

Let us consider a first possible realization of the "small-size" $\sigma
-$ordering regime. This is obtained assuming that the diameter of the
hard-spheres $\sigma$ is considered $\ll1$ in a suitable sense, while the
scale-length $L$ is ordered according to the prescriptions that $\sigma/L\sim
O(\varepsilon^{\frac{1}{2}})$ and letting
\begin{equation}
\left\{
\begin{array}
[c]{c}%
\sigma\sim O(\varepsilon^{\alpha})\\
L\sim O(\varepsilon^{\alpha-\frac{1}{2}})\\
L\sim L_{o}\sim L_{\rho}\\
\alpha\in\left]  0,\frac{1}{3}\right]  .
\end{array}
\right.  \label{dilute-gas ordering regime}%
\end{equation}
Introducing the dimensionless parameter
\begin{equation}
\Delta_{L}\equiv\frac{4\pi N\sigma^{3}}{3L^{3}} \label{delta-L}%
\end{equation}
this implies necessarily that $\Delta_{L}\sim O(\varepsilon^{\frac{1}{2}})$
and hence, due to the inequalities $L\leq L_{o},$ and $\Delta\leq\Delta_{L}$
also $\Delta\lesssim O(\varepsilon^{\frac{1}{2}})$. Therefore the orderings
(\ref{dilute-gas ordering regime}) necessarily equivalently identify a
dilute-gas ordering\emph{,} which \ can therefore be characterized as a
\emph{collisional, dilute-gas and small-size }$\sigma-$\emph{ordering regime.
}In particular, when $\alpha=\frac{1}{2}$ and $L\sim L_{o}$ it follows that
$\Delta_{L}\sim\Delta\sim O(\varepsilon^{\frac{1}{2}})$ and $K_{n}\sim
O(\varepsilon^{0}),$ so that the customary dilute-gas ordering considered
originally by Grad \cite{Grad} (see also Refs. \cite{noi3,noi7}) is recovered.
This is obtained requiring%

\begin{equation}
\left\{
\begin{array}
[c]{c}%
N\sigma^{2}\sim O(\varepsilon^{0})\\
L\sim L_{o}\sim O(\varepsilon^{0}).
\end{array}
\right.  \label{GRAD-DILUTE-GAS-ORD}%
\end{equation}

\subsection{3B - Dilute-gas finite-size $\sigma-$ordering regime}

Another type of ordering regime is obtained requiring $\sigma$ to be finite
while prescribing again the scale-length $L$ in such a way to satisfy the
requirement $K_{n}\sim O(\varepsilon^{0}).$ Let us require%

\begin{equation}
\left\{
\begin{array}
[c]{c}%
\sigma\sim O(\varepsilon^{0})\\
L\sim O(\varepsilon^{-\frac{1}{2}})\\
L\sim L_{\rho}\lesssim L_{o}.
\end{array}
\right.  \label{finite-size-dilute-gas}%
\end{equation}
Notice that again $\Delta_{L}\sim O(\varepsilon^{\frac{1}{2}})$ and hence
$\Delta\lesssim O(\varepsilon^{\frac{1}{2}}).$ Therefore the ordering
(\ref{finite-size-dilute-gas}) corresponds to a dilute-gas\emph{ }ordering
which will be referred to as \emph{collisional, dilute-gas and finite-size
}$\sigma-$\emph{ordering regime. }\bigskip

Finally, for completeness we point out possible realizations of dense-gas
ordering regimes.

\subsection{3C - Dense-gas ordering regimes}

Let us require for this purpose that the parameter $\Delta$ is of order
$O(\varepsilon^{0}),$\textit{ i.e.,} that the hard-sphere system is dense.
This happens in the case in which $\sigma/L_{o}\sim O(\varepsilon^{\frac{1}%
{3}}).$ Since by construction due to the inequality $L\leq L_{o}$ also
$\Delta\leq\Delta_{L}$ it follows necessarily that it must be $\Delta
\sim\Delta_{L}$ and hence $L\sim L_{o}$ too. This means that $K_{n}\sim
O(\varepsilon^{-\frac{1}{3}})$ which therefore corresponds necessarily to a
strongly-collisional regime. Let us assume for this purpose that the following
orderings apply:%
\begin{equation}
\left\{
\begin{array}
[c]{c}%
\sigma\sim O(\varepsilon^{\alpha})\\
L\sim O(\varepsilon^{\alpha-\frac{1}{3}})\\
L\sim L_{\rho}\sim L_{o}.\\
\alpha\in\left[  0,\infty\right]  .
\end{array}
\right.  \label{small-size-strongly collisional}%
\end{equation}
Therefore this implies that necessarily $\Delta\sim\Delta_{L}\sim
O(\varepsilon^{0})$. In particular, if $\alpha>0$ this corresponds to a
small-size $\sigma-$ordering regime, to be referred to as
\emph{strongly-collisional, dense-gas and small-size }$\sigma-$\emph{ordering
regime.} A special case is provided by the choice $\alpha=0,$ which
corresponds instead to a finite-size $\sigma-$ordering regime for which \emph{
}%
\begin{equation}
\left\{
\begin{array}
[c]{c}%
\sigma\sim O(\varepsilon^{0})\\
L\sim O(\varepsilon^{-\frac{1}{3}}),\\
L\sim L_{\rho}\sim L_{o}.
\end{array}
\right.  \label{finite-size strongly collisional}%
\end{equation}
This can therefore be characterized as \emph{strongly-collisional, dense-gas
and finite-size }$\sigma-$\emph{ordering.}

\bigskip

\section{4 - Asymptotic approximations of the Master kinetic equation}

In this section we intend to pose the problem of the construction of the
asymptotic approximations of the Master kinetic equation which are appropriate
for the treatment of most of the asymptotic regimes discussed in subsections
3A, 3B and 3C. In detail we intend to show that:

\begin{itemize}
\item \emph{First asymptotic approximation}: in the ordering regime
(\ref{dilute-gas ordering regime}) to lowest order in $O(\varepsilon)$ the
Master equation reduces to the Boltzmann kinetic equation, with the Master
collision operator being approximated in this case by the collision operator%
\begin{gather}
\overline{C}_{1MB}(\left.  \overline{\rho}_{1}^{(N)}\right\vert \overline
{\rho}_{1}^{(N)})=\frac{N\sigma}{L^{2}}^{2}\int\limits_{U_{1(2)}}%
d\overline{\mathbf{v}}_{2}\int^{(+)}d\Sigma_{21}f(\mathbf{r}_{1}%
,\mathbf{r}_{2}=\mathbf{r}_{1},t).\nonumber\\
\left[  \overline{\rho}_{1}^{(N)}(\mathbf{r}_{1},\mathbf{v}_{1}^{(.-)}%
,t)\overline{\rho}_{1}^{(N)}(\mathbf{r}_{1},\mathbf{v}_{2}^{(-)}%
,t)-\overline{\rho}_{1}^{(N)}(\mathbf{r}_{1},\mathbf{v}_{1},t)\overline{\rho
}_{1}^{(N)}(\mathbf{r}_{1},\mathbf{v}_{2},t)\right]  \left\vert \overline
{\mathbf{v}}_{21}\cdot\mathbf{n}_{21}\right\vert
,\label{ASYMP-APPROXIMATION-1}%
\end{gather}
where $f(\mathbf{r}_{1},\mathbf{r}_{2},t)$\textbf{ }is the strictly-positive
weight-factor prescribed by Eq. (\ref{OCCUPATION COEFFICIENTS TERM}). In the
expression of same equation the occupation coefficients are here approximated
as follows:%
\begin{equation}
\left\{
\begin{array}
[c]{c}%
k_{1}^{(N)}(\mathbf{r}_{1},t)\cong1-\frac{N}{2}\int\limits_{\Phi_{12}%
}d\mathbf{r}_{2}n_{1}^{(N)}(\mathbf{r}_{2},t),\\
k_{2}^{(N)}(\mathbf{r}_{1},\mathbf{r}_{2},t)\cong1-\frac{3N}{4}\int
\limits_{\Phi_{12}}d\mathbf{r}_{2}n_{1}^{(N)}(\mathbf{r}_{2},t)-\frac{3N}%
{4}\int\limits_{\Phi_{23}}d\mathbf{r}_{3}n_{1}^{(N)}(\mathbf{r}_{3},t),
\end{array}
\right.  \label{OCC-1-ASYMP}%
\end{equation}
with $\Phi_{ij}$ denoting the hard-sphere interior domain $\Phi_{ij}%
\equiv\left\{  \mathbf{r}_{j}:\left\vert \mathbf{r}_{j}-\mathbf{r}%
_{i}\right\vert <\sigma\right\}  .$ This yields therefore the asymptotic
approximation%
\begin{equation}
f(\mathbf{r}_{1},\mathbf{r}_{2},t)\cong\frac{1-\frac{3N}{4}\int\limits_{\Phi
_{12}}d\mathbf{r}_{2}n_{1}^{(N)}(\mathbf{r}_{2},t)-\frac{3N}{4}\int
\limits_{\Phi_{23}}d\mathbf{r}_{3}n_{1}^{(N)}(\mathbf{r}_{3},t)}{1-\frac{N}%
{2}\int\limits_{\Phi_{12}}d\mathbf{r}_{2}n_{1}^{(N)}(\mathbf{r}_{2}%
,t)-\frac{N}{2}\int\limits_{\Phi_{23}}d\mathbf{r}_{3}n_{1}^{(N)}%
(\mathbf{r}_{3},t)}.\label{ASYMP-APPROX}%
\end{equation}

The following remarks are in order regarding the collision operator
$\overline{C}_{1MB}(\left.  \overline{\rho}_{1}^{(N)}\right\vert
\overline{\rho}_{1}^{(N)})$. \ First one notices that it provides a
generalization of the Boltzmann collision operator (issue \#2)$.$ In
particular, one can readily show (see the proof reported below) that to order
$O(\varepsilon^{0})$ it coincides by construction with the customary Boltzmann
collision operator, since then the weight-factor\textbf{ }$f(\mathbf{r}%
_{1},\mathbf{r}_{2},t)$ can be approximated with unity. The asymptotic
approximate formula (\ref{ASYMP-APPROX}) for the weight-factor\textbf{
}$f(\mathbf{r}_{1},\mathbf{r}_{2},t)$\textbf{ }given by Eq.
(\ref{OCCUPATION COEFFICIENTS TERM}) retains, instead, also leading-order
corrections which are produced by the $1-$ and $2-$body occupation
coefficients (issue \#3). \ Second, the structure of the collision operator
(\ref{ASYMP-APPROXIMATION-1}) has also formal analogies with the one
introduced by Enskog in his namesake equation. The key feature in this case
lies in the prescription of the weight-factor $f(r_{1},r_{2},t)$ which is here
provided by Eq. (\ref{ASYMP-APPROX}) while remaining in principle undetermined
in the context of the Enskog kinetic equation. Thus, provided, the same
prescription indicated above is made for\textbf{ }$f(\mathbf{r}_{1}%
,\mathbf{r}_{2},t)$, the collision operator (\ref{ASYMP-APPROXIMATION-1}) can
be viewed as realizing also an approximate representation of the Enskog
collision operator (issue \#4).

\item \emph{Second asymptotic approximation}: in\ validity of the ordering
regimes (\ref{finite-size-dilute-gas}) the Master equation reduces, instead,
to an asymptotic Master kinetic equation determined by the collision operator%
\begin{gather}
\overline{C}_{1}(\left.  \overline{\rho}_{1}^{(N)}\right\vert \overline{\rho
}_{1}^{(N)})=\frac{N\sigma}{L^{2}}^{2}\int\limits_{U_{1(2)}}d\overline
{\mathbf{v}}_{2}\int^{(-)}d\Sigma_{21}f(\mathbf{r}_{1},\mathbf{r}%
_{2},t).\nonumber\\
\left[  \overline{\rho}_{1}^{(N)}(\mathbf{r}_{1},\mathbf{v}_{1}^{(.-)}%
,t)\overline{\rho}_{1}^{(N)}(\mathbf{r}_{2},\mathbf{v}_{2}^{(-)}%
,t)-\overline{\rho}_{1}^{(N)}(\mathbf{r}_{1},\mathbf{v}_{1},t)\overline{\rho
}_{1}^{(N)}(\mathbf{r}_{2},\mathbf{v}_{2},t)\right]  \left\vert \overline
{\mathbf{v}}_{21}\cdot\mathbf{n}_{21}\right\vert \overline{\Theta}^{\ast
}\label{ASYMP-APPROXIMATION-2}%
\end{gather}
in which the weight-factor $f(\mathbf{r}_{1},\mathbf{r}_{2},t)$ is expressed
in terms of the asymptotic estimate (\ref{ASYMP-APPROX}) and due to the
requirement that $\sigma$ remains finite, so that necessarily $\mathbf{r}%
_{2}=\mathbf{r}_{1}+\sigma\mathbf{n}_{21}$. This implies that although Eq.
(\ref{ASYMP-APPROXIMATION-2}) has formal analogies with the customary form of
the Enskog collision operator, two major differences arise. The first one lies
in the prescription of the weight-factor itself, which in the present case is
determined by Eq. (\ref{ASYMP-APPROX}) while it choice remains unspecified in
the context of the Enskog statistical approach. The second follows because of
the adoption of MCBC requiring that in Eqs. (\ref{ASYMP-APPROXIMATION-2}) the
solid-angle integration must be carried out on the subset $\int^{(-)}%
d\Sigma_{21}$ of\textbf{ }incoming particles for which\textbf{ }%
$\mathbf{v}_{12}\cdot\mathbf{n}_{12}<0$\textbf{ }instead on the complementary
set $\int^{(-)}d\Sigma_{21}$ as done in the Enskog collision operator (issue \#4).

\item \emph{Third asymptotic approximation}: in the ordering regime
(\ref{small-size-strongly collisional}) subject to the requirement $\alpha>0$
to leading order in $O(\varepsilon)$ the Master equation reduces to the
asymptotic Master kinetic equation, expressed in terms of the collision
operator which to leading order in $O(\varepsilon)$ reads%
\begin{gather}
\overline{C}_{1}(\left.  \overline{\rho}_{1}^{(N)}\right\vert \overline{\rho
}_{1}^{(N)})=\frac{N\sigma}{L^{2}}^{2}\int\limits_{U_{1(2)}}d\overline
{\mathbf{v}}_{2}\int^{(+)}d\Sigma_{21}f(\mathbf{r}_{1},\mathbf{r}%
_{2}=\mathbf{r}_{1},t)\times\nonumber\\
\left[  \overline{\rho}_{1}^{(N)}(\mathbf{r}_{1},\mathbf{v}_{1}^{(.-)}%
,t)\overline{\rho}_{1}^{(N)}(\mathbf{r}_{1},\mathbf{v}_{2}^{(-)}%
,t)-\overline{\rho}_{1}^{(N)}(\mathbf{r}_{1},\mathbf{v}_{1},t)\overline{\rho
}_{1}^{(N)}(\mathbf{r}_{1},\mathbf{v}_{2},t)\right]  \left\vert \overline
{\mathbf{v}}_{21}\cdot\mathbf{n}_{21}\right\vert
,\label{ASYMP-APPROXIMATION-3}%
\end{gather}
where $f(\mathbf{r}_{1},\mathbf{r}_{2},t)$ is prescribed again by Eq.
(\ref{OCCUPATION COEFFICIENTS TERM}). However, now in difference with the two
cases indicated above the asymptotic approximations (\ref{OCC-1-ASYMP}) and
(\ref{ASYMP-APPROX}) do not hold, so that the occupation coefficients
$k_{1}^{(N)}(\mathbf{r}_{1},t)$ and $k_{2}^{(N)}(\mathbf{r}_{1},\mathbf{r}%
_{2},t)$ need to be determined iteratively in terms of Eqs. (\ref{OCC-1}).
\end{itemize}

Finally, we mention that the case represented by the ordering
(\ref{finite-size strongly collisional}) must be treated separately, in the
sense that no approximation is actually possible on the functional form of the
Master collision operator (\ref{MASTER-coll}).

\subsection{4A - Proof of the first asymptotic approximation}

Let us first prove the validity of Eqs. (\ref{ASYMP-APPROXIMATION-1}) and
(\ref{ASYMP-APPROX}). For this purpose one first notices that thanks to the
ordering regime (\ref{dilute-gas ordering regime}) the $1-$body PDFs $\rho
_{1}^{(N)}(\mathbf{r}_{2},\mathbf{v}_{2}^{(+)},t)$ and $\rho_{1}%
^{(N)}(\mathbf{r}_{2},\mathbf{v}_{2},t)$ can be approximated in terms of
$\rho_{1}^{(N)}(\mathbf{r}_{1},\mathbf{v}_{2}^{(+)},t)$ and $\rho_{1}%
^{(N)}(\mathbf{r}_{1},\mathbf{v}_{2},t)$ respectively. For the same reason the
occupation coefficients $k_{1}^{(N)}(\mathbf{r}_{2},t)$ and $k_{2}%
^{(N)}(\mathbf{r}_{1},\mathbf{r}_{2},t)$ can be approximated in terms of
$k_{1}^{(N)}(\mathbf{r}_{1},t)$ and $k_{2}^{(N)}(\mathbf{r}_{1},\mathbf{r}%
_{2}\equiv\mathbf{r}_{1},t)$. Second, again thanks to
Eqs.(\ref{dilute-gas ordering regime}),\ in the Master collision operator the
solid-angle integration on the sub-domain $\mathbf{v}_{12}\cdot\mathbf{n}%
_{12}<0$ (namely $\int^{(-)}d\Sigma_{21}$) can be equivalently exchanged with
the corresponding complementary subset $\mathbf{v}_{12}\cdot\mathbf{n}%
_{12}\geq0,$\textit{ i.e.,} $\int^{(+)}d\Sigma_{21},$ while the domain theta
function $\overline{\Theta}^{\ast}(\mathbf{x}_{2})$ becomes $\overline{\Theta
}^{\ast}(\mathbf{x}_{2})\equiv\overline{\Theta}\left(  \left\vert
\mathbf{r}_{1}\right\vert \right)  $ so that its contribution to the collision
integral is ignorable. Third, to prove the asymptotic estimate
(\ref{ASYMP-APPROX}), let us notice that in validity of the ordering
(\ref{dilute-gas ordering regime}) it follows that $\int\limits_{\Phi_{12}%
}d\overline{\mathbf{r}}_{2}\overline{n}_{1}^{(N)}(\mathbf{r}_{2},t)\sim$
$\overline{n}_{1}^{(N)}(\mathbf{r}_{2}^{\ast},t)\frac{4\pi\sigma^{3}}{3L^{3}%
}\equiv\frac{\sigma^{3}}{L^{3}}\widehat{x}$, with $\widehat{x}$ a suitable
mean value such that $\widehat{x}\sim O(\varepsilon^{0}).$ As a consequence
from Eqs.(\ref{OCC-1-ASYMP}) in order of magnitude it follows that%
\begin{equation}
\left\{
\begin{array}
[c]{c}%
k_{1}^{(N)}(\mathbf{r}_{1},t)\sim1-\frac{N\sigma^{3}\widehat{x}}{2L^{3}}\\
k_{2}^{(N)}(\mathbf{r}_{1},\mathbf{r}_{2},t)\sim1-\frac{3N\sigma^{3}%
\widehat{x}}{2L^{3}}%
\end{array}
\right.  \label{OCC-1-ASYMP-BIS}%
\end{equation}
which implies%
\begin{equation}
f(\mathbf{r}_{1},\mathbf{r}_{2},t)\sim1-\frac{N\sigma^{3}\widehat{x}}{2L^{3}%
}.\label{ASTYMP-APPROX}%
\end{equation}
The proof of the asymptotic estimates (\ref{OCC-1-ASYMP-BIS}) is
straightforward. In fact, Eq.(\ref{OCC-1}) then requires, based on the
mean-value theorem, implies that%
\begin{align}
\left(  k_{1}^{(N)}(\mathbf{r}_{1}^{\ast},t)\right)  ^{N} &  \sim
\int\limits_{\Gamma_{1(2)}}d\overline{\mathbf{x}}_{2}\overline{\rho}_{1}%
^{(N)}(\mathbf{x}_{2},t)\int\limits_{\overline{\Gamma}_{1(3)}^{(1)}}%
d\overline{\mathbf{x}}_{3}\overline{\rho}_{1}^{(N)}(\mathbf{x}_{3}%
,t)\int\limits_{\overline{\Gamma}_{1(N)}^{(1c)}}d\overline{\mathbf{x}}%
_{N}\overline{\rho}_{1}^{(N)}(\mathbf{x}_{N},t)\equiv\nonumber\\
&  \equiv\int\limits_{\overline{\Omega}_{1(2)}}d\mathbf{r}_{2}n_{1}%
^{(N)}(\mathbf{r}_{2},t)\int\limits_{\overline{\Omega}_{1(3)}^{(1)}%
}d\mathbf{r}_{3}n_{1}^{(N)}(\mathbf{r}_{3},t)\int\limits_{\overline{\Omega
}_{1(N)}^{(1c)}}d\mathbf{r}_{N}n_{1}^{(N)}(\mathbf{r}_{N},t),\label{Previous}%
\end{align}
with $k_{1}^{(N)}(\mathbf{r}_{1}^{\ast},t)$ a suitable mean-value. Therefore
the same equation yields the asymptotic estimate%
\begin{equation}
\left(  k_{1}^{(N)}(\mathbf{r}_{1}^{\ast},t)\right)  ^{N}\sim\left(
1-\frac{\sigma^{3}\widehat{x}}{_{L^{3}}}\right)  \left(  1-2\frac{\sigma
^{3}\widehat{x}}{_{L^{3}}}\right)  ...\left(  1-(N-1)\frac{\sigma^{3}%
\widehat{x}}{_{L^{3}}}\right)
\end{equation}
which is manifestly consistent with Eq.(\ref{OCC-1-ASYMP-BIS}). The proof of
the asymptotic estimates for $k_{2}^{(N)}(\mathbf{r}_{1},\mathbf{r}_{2},t)$
and (\ref{ASTYMP-APPROX}) is analogous, thus yielding the consistency of the
asymptotic approximations (\ref{OCC-1-ASYMP}) and (\ref{ASYMP-APPROX}).

\subsection{4B - Proof of the second asymptotic approximation}

The proof of Eq.(\ref{ASYMP-APPROXIMATION-2}) is similar as far as the
asymptotic estimate (\ref{ASTYMP-APPROX}) is concerned. Now, however, due to
the finite size of the hard spheres (see Eqs. (\ref{finite-size-dilute-gas}))
the correct spatial dependences must be retained in the $1-$body PDF's
$\rho_{1}^{(N)}(\mathbf{r}_{2},\mathbf{v}_{2}^{(+)},t)$ and $\rho_{1}%
^{(N)}(\mathbf{r}_{2},\mathbf{v}_{2},t)$ which must both be evaluated at the
position $\mathbf{r}_{2}=\mathbf{r}_{1}+\sigma\mathbf{n}_{21}.$ As a
consequence the corresponding asymptotic approximation
(\ref{ASYMP-APPROXIMATION-2}) manifestly holds for the Master collision operator.

\subsection{4C - Proof of the third asymptotic approximation}

The proof of Eq. (\ref{ASYMP-APPROXIMATION-3}) is similarly straightforward.
In fact, first one notices that in close analogy with case 4A, thanks to the
small-size assumption introduced for $\sigma$ , the $1-$body PDFs $\rho
_{1}^{(N)}(\mathbf{r}_{2},\mathbf{v}_{2}^{(+)},t),$ $\rho_{1}^{(N)}%
(\mathbf{r}_{2},\mathbf{v}_{2},t)$ as well as the occupation coefficients
$k_{1}^{(N)}(\mathbf{r}_{2},t)$ and $k_{2}^{(N)}(\mathbf{r}_{1},\mathbf{r}%
_{2},t)$ can all be approximated replacing $\mathbf{r}_{2}\rightarrow
\mathbf{r}_{1}.$ As a consequence again the solid-angle integration
$\int^{(-)}d\Sigma_{21}$ can be equivalently be evaluated in terms of the
outgoing-particle subset $\int^{(+)}d\Sigma_{21}$ while the contribution of
the theta function $\overline{\Theta}^{\ast}$ is ignorable. \ Finally, due to
the dense-gas asymptotic ordering included in
(\ref{small-size-strongly collisional}) no obvious asymptotic approximation is
available for the occupation coefficients $k_{1}^{(N)}(\mathbf{r}_{1},t)$ and
$k_{2}^{(N)}(\mathbf{r}_{1},\mathbf{r}_{2}=\mathbf{r}_{1},t)$. \ Therefore
their exact expression following from Eqs. (\ref{OCC-1}) must be retained in
Eq. (\ref{ASYMP-APPROXIMATION-3}).

\section{5 - Conclusions}

In this paper the problem has been addressed of identifying possible
physically-meaningful asymptotic approximations of the Master kinetic equation
which apply to ,\textit{ i.e.,} formed by $N\equiv\frac{1}{\varepsilon}$%
\ $\gg1$ hard-spheres. The statistical approach has been based on the
"\textit{ab initio}" axiomatic statistical theory recently developed
\cite{noi1,noi2,noi3,noi4,noi5,noi6,noi7,noi8,noi9}. As a result, once the
Master kinetic equation is cast in dimensionless form, the existence of
multiple asymptotic ordering regimes for the same equation has been pointed
out which hold for large $N-$body systems. These regimes correspond to
appropriate prescriptions of the relevant physical parameters of the same
equation and include, as a particular possible realization, the customary
dilute-gas ordering originally introduced by Grad \cite{Grad} for his
construction of the Boltzmann kinetic equation. The new ordering regimes
encompass either small or finite-size hard-spheres as well as dilute or dense,
collisional or strongly-collisional particle systems. In particular possible
realizations include:

\begin{itemize}
\item \emph{the dilute-gas small-size }$\sigma-$\emph{ordering regime
}(prescribed by the ordering Eqs. (\ref{dilute-gas ordering regime}));

\item \emph{the dilute-gas finite-size }$\sigma-$\emph{ordering regime }(in
the sense of Eqs. (\ref{finite-size-dilute-gas}));

\item \emph{the dense-gas ordering regime} (see Eqs.\emph{ }%
(\ref{small-size-strongly collisional}) in the case in which $\alpha>0$).
\end{itemize}

\noindent Corresponding asymptotic approximations have been determined for the
Master collision operator, displaying also their relationship/difference with
respect to the Boltzmann and Enskog collision operators.

The present results are believed to be crucial both in kinetic theory and
fluid dynamics.\ Indeed, regarding possible challenging future developments
one should particularly mention possible applications both of the Master
kinetic equation itself as well as of the asymptotic approximations here
pointed out for the first time. The hard-sphere kinetic statistical treatment
based on these equations is expected to successfully apply to a variety of
complex fluid-dynamics systems as well as to neutral and/or ionized gases of
interest for laboratory research and astrophysics.

\section{Acknowledgments}

Work developed within the research projects: A) the Albert Einstein Center for
Gravitation and Astrophysics, Czech Science Foundation No. 14-37086G; B) the
grant No. 02494/2013/RRC \textquotedblleft\textit{kinetick\'{y}
p\v{r}\'{\i}stup k proud\u{e}n\'{\i} tekutin}\textquotedblright\ (kinetic
approach to fluid flow) in the framework of the \textquotedblleft Research and
Development Support in Moravian-Silesian Region\textquotedblright, Czech
Republic: C) the research projects of the Czech Science Foundation GA\v{C}R
grant No. 14-07753P. Initial framework and motivations of the investigation
were based on the research projects developed by the Consortium for
Magnetofluid Dynamics (University of Trieste, Italy) and, in reference to
issues \#2 and \#3, the MIUR (Italian Ministry for Universities and Research)
PRIN Research Program \textquotedblleft\textit{Problemi Matematici delle
Teorie Cinetiche e Applicazioni}\textquotedblright, University of Trieste,
Italy. The authors are grateful to the International Center for Theoretical
Physics (Miramare, Trieste, Italy) for the hospitality during the preparation
of the manuscript.

\end{document}